\documentclass[aps,pra,reprint,superscriptaddress]{revtex4-1}
\usepackage{amsmath}
\usepackage{graphicx}
\usepackage{longtable}
\usepackage{hyperref}
\usepackage{amssymb}
\usepackage{dcolumn}
\usepackage{mhchem}
\begin{document}

\title{Asymptotic Correction Schemes for Semilocal Exchange-Correlation Functionals} 

\author{Chi-Ruei Pan} 
\affiliation{Department of Physics, National Taiwan University, Taipei 10617, Taiwan} 

\author{Po-Tung Fang} 
\affiliation{Department of Physics, National Taiwan University, Taipei 10617, Taiwan} 

\author{Jeng-Da Chai} 
\email[Author to whom correspondence should be addressed. Electronic mail: ]{jdchai@phys.ntu.edu.tw.} 
\affiliation{Department of Physics, National Taiwan University, Taipei 10617, Taiwan} 
\affiliation{Center for Theoretical Sciences and Center for Quantum Science and Engineering, National Taiwan University, Taipei 10617, Taiwan} 

\date{\today}

\begin{abstract} 

Aiming to remedy the incorrect asymptotic behavior of conventional semilocal exchange-correlation (XC) density functionals for finite systems, we propose an asymptotic correction scheme, wherein an exchange 
density functional whose functional derivative has the correct $(-1/r)$ asymptote can be directly added to any semilocal density functional. In contrast to semilocal approximations, our resulting exchange kernel 
in reciprocal space exhibits the desirable singularity of the type $O(-1/q^2)$ as $q \rightarrow 0$, which is a necessary feature for describing the excitonic effects in non-metallic solids. By applying this scheme to a 
popular semilocal density functional, PBE [J. P. Perdew, K. Burke, and M. Ernzerhof, Phys. Rev. Lett. {\bf 77}, 3865 (1996)], the predictions of the properties that are sensitive to the asymptote are significantly improved, 
while the predictions of the properties that are insensitive to the asymptote remain essentially the same as PBE. Relative to the popular model XC potential scheme, our scheme is significantly superior for 
ground-state energies and related properties. In addition, without loss of accuracy, two closely related schemes are developed for the efficient treatment of large systems. 

\end{abstract}

\maketitle

\section{Introduction}

Over the past two decades, Kohn-Sham density functional theory (KS-DFT) \cite{HK,KSDFT} has been one of the most powerful theoretical methods for the ground-state properties of large electronic systems. 
Its time-dependent extension, time-dependent density functional theory (TDDFT) \cite{TDDFT1,TDDFT2,TDDFT3} has gradually become popular for the study of excited-state and time-dependent properties. 

In KS-DFT, the exact exchange-correlation (XC) density functional $E_{xc}[\rho]$ remains unknown and needs to be approximated. Accurate density functional approximations to $E_{xc}[\rho]$ have been 
successively developed to extend the applicability of KS-DFT to a wide variety of systems. Despite the recent advances in the orbital-dependent density functional approach \cite{OD}, semilocal density functionals 
remain popular due to their computational efficiency for large systems and reasonable accuracy for applications governed by short-range XC effects \cite{semilocal}. However, due to the associated several 
qualitative failures, semilocal functionals can produce erroneous results in situations where the accurate treatment of nonlocality of the XC hole is important \cite{SciYang,SciYang2,SciChai}. 

One of the important and long-standing subjects in KS-DFT is the asymptotic behavior of the XC potential $v_{xc}(\textbf{r}) = \delta E_{xc}[\rho]/\delta \rho(\textbf{r})$. 
For finite systems, the exact $v_{xc}(\textbf{r})$ exhibits the Coulombic $(-1/r)$ decay as $r \rightarrow \infty$ \cite{exactip1,exactip2,sum_vxc2a,sum_vxc2b}. 
However, due to the severe self-interaction error (SIE) \cite{SIC}, the XC potential of semilocal functionals fails to describe the correct asymptotic behavior, yielding qualitatively incorrect predictions for 
the properties sensitive to the asymptote, such as the vertical ionization potentials and high-lying (Rydberg) excitation energies of atoms and molecules \cite{HOMO_BAD1,HOMO_BAD2}. 

Currently, perhaps the most successful density functional methods in practice to improve the asymptote of the XC potential are provided by the long-range corrected (LC) hybrid 
scheme \cite{LC-DFT,LCHirao,CAM-B3LYP,LC-wPBE,BNL,wB97X,wB97X-D,op,wB97X-2,wM05-D,LC-D3} and asymptotically corrected (AC) model potential scheme \cite{LB94,AC0,AC1,LBa,AC2,AC3,AC4,AC5,AC6,AC7,AC8}. 
For the LC hybrid scheme, the nonlocal Hartree-Fock (HF) exchange for the long-range electron-electron interactions is added to a semilocal functional. Therefore, the LC hybrid scheme can be impractical for very large systems 
due to the inclusion of the long-range HF exchange (which significantly increases the computational cost relative to the semilocal functional). By contrast, for the AC model potential scheme, an AC XC potential is directly modeled 
and added to a semilocal functional, maintaining the similar cost as the semilocal functional. In principle, a model XC potential should be a functional derivative of some $E_{xc}[\rho]$. 
However, as a number of popular model potentials are found {\it not} to be functional derivatives, several necessary conditions for a functional derivative can be violated \cite{Staroverov}. Besides, as these model potentials 
are not variationally stable, the associated XC energies are not uniquely defined, and properties obtained from these model potentials need to be carefully interpreted \cite{Staroverov,Staroverov2b}. 
Recently, we have examined the performance of the LC hybrid scheme and AC model potential scheme on a very wide range of applications \cite{LCAC}. 
Despite its computational efficiency, the popular model potential scheme can exhibit severe errors in the calculated ground-state energies and related properties, due to the lack of $E_{xc}[\rho]$. 

On the other hand, for a system of $N$ electrons, the Fermi-Amaldi (FA) XC functional \cite{FA}, 
\begin{equation}
\label{eq:FA}
E^{\text{FA}}_{xc}[\rho] = -\frac{1}{2N} \iint \frac{\rho(\textbf{r})\rho(\textbf{r}')}{\left|{\bf r} - {\bf r}' \right|} d\textbf{r}d\textbf{r}', 
\end{equation} 
which is simply $(-1/N)$ times the Hartree energy functional, appears to be the simplest XC functional whose functional derivative has the correct $(-1/r)$ asymptote. However, there are several problems with 
the FA model \cite{FA_size}. While the FA XC potential is correct in the asymptotic region, it is inaccurate elsewhere. Besides, due to its delocalized XC hole, the FA model is not size-consistent, where the energy of 
a system composed of two or more well-separated subsystems is {\it not} identical to the sum of the energies of the separate subsystems \cite{separability}. 

In this work, we propose an AC scheme for any system composed of atoms (e.g., atoms, molecules, and solids), wherein a modified FA XC functional, which is size-consistent in the calculated energy 
and whose functional derivative has the correct $(-1/r)$ asymptote, can be directly added to any semilocal functional. Without loss of accuracy, two related efficient schemes are also developed for large systems.

\section{Theoretical Methods}

\subsection{LFA scheme}

By partitioning and localizing a modified FA XC hole to the atoms in a system, we propose the ``localized" FA (LFA) exchange functional, 
\begin{equation}
\label{eq:9}
\begin{split}
&E^{\text{LFA}}_{x}[\rho_{\alpha},\rho_{\beta}] = -\sum_{\sigma=\alpha,\beta}\sum_{A} \frac{1}{2N_{A,\sigma}} \\ 
& \times \iint \rho_{A,\sigma}({\bf r}) \rho_{A,\sigma}({\bf r}') \frac{\text{erf}(\omega \left|{\bf r} - {\bf r}' \right|)}{\left|{\bf r} - {\bf r}' \right|} d\textbf{r}d\textbf{r}', 
\end{split}
\end{equation}
to resolve the size-inconsistency issue associated with the FA model \cite{Sup}. Here, the second sum is over all the atoms in the system, $\rho_{A,\sigma}(\textbf{r})$ is the $\sigma$-spin ($\sigma$ = $\alpha$ 
for spin up or $\beta$ for spin down) electron density associated with the atom $A$, 
\begin{equation}
\label{eq:10} 
\rho_{A,\sigma}(\textbf{r}) = w_{A}(\textbf{r}) \rho_{\sigma}(\textbf{r}),
\end{equation} 
and the weight function $w_{A}(\textbf{r})$, ranging between 0 and 1, is of the Hirshfeld type \cite{AIM1,AIM2}: 
\begin{equation}
\label{eq:11} 
w_{A}(\textbf{r}) = \frac{\rho^{0}_{A}(\textbf{r})}{\sum_{B} \rho^{0}_{B}(\textbf{r})}, 
\end{equation}
where $\rho^{0}_{A}(\textbf{r})$ is the spherically averaged electron density computed for the isolated atom $A$. $N_{A,\sigma}$ is the number of the $\sigma$-spin electrons associated with the atom $A$, 
\begin{equation}
\label{eq:NA} 
N_{A,\sigma} = \int \rho_{A,\sigma}(\textbf{r}) d\textbf{r}, 
\end{equation} 
and the long-range interelectron repulsion operator $\text{erf}(\omega \left|{\bf r} - {\bf r}' \right|)/\left|{\bf r} - {\bf r}' \right|$ is to retain the correct asymptotic behavior without the (unneeded) energy contribution 
from the complementary short-range operator, where $\omega$ is a parameter defining the range of the operators. Due to the sum rule of $\sum_{A} w_{A}(\textbf{r}) = 1$, 
$\sum_{A} \rho_{A,\sigma}(\textbf{r})$ = $\rho_{\sigma}(\textbf{r})$ and $\sum_{A} N_{A,\sigma} = N_{\sigma}$ (the number of $\sigma$-spin electrons). 

By taking the functional derivative of $E^{\text{LFA}}_{x}[\rho_{\alpha},\rho_{\beta}]$, the LFA exchange potential for $\sigma$-spin electrons is 
\begin{equation}
\label{eq:potential}
\begin{split}
v^{\text{LFA}}_{x,\sigma}(\textbf{r}) & = \frac{\delta E^{\text{LFA}}_{x}[\rho_{\alpha},\rho_{\beta}]} {\delta\rho_{\sigma}({\bf r})} \\ 
& = -\sum_{A} \frac{w_{A}(\textbf{r})}{N_{A,\sigma}} \int \rho_{A,\sigma}(\textbf{r}') \frac{\text{erf}(\omega \left|{\bf r}-{\bf r}'\right|)}{\left|{\bf r}-{\bf r}'\right|} d\textbf{r}'. 
\end{split}
\end{equation} 
If the functional derivative of $N_{A,\sigma}$ is also taken, an additional constant term 
\begin{equation}
\begin{split}
v^{\text{LFA}}_{x,\sigma}(\infty) & \equiv \sum_{A} \frac{1}{2N_{A,\sigma}^2} \\
& \times \iint \rho_{A,\sigma}({\bf r}) \rho_{A,\sigma}({\bf r}') \frac{\text{erf}(\omega \left|{\bf r} - {\bf r}' \right|)}{\left|{\bf r} - {\bf r}' \right|} d\textbf{r}d\textbf{r}' 
\end{split}
\end{equation} 
should be added to Eq.\ (\ref{eq:potential}). However, as will be shown later, this constant is of no consequence. 

In the asymptotic limit, $v^{\text{LFA}}_{x,\sigma}(\textbf{r})$ has the correct asymptotic form, 
\begin{equation}
\label{eq:potential2}
\begin{split}
\lim_{r \rightarrow \infty} v^{\text{LFA}}_{x,\sigma}(\textbf{r}) & = -\sum_{A} \frac{w_{A}(\textbf{r})}{N_{A,\sigma}} \int \rho_{A,\sigma}(\textbf{r}') \frac{1}{\left|{\bf r}\right|} d\textbf{r}' \\ 
& = -\frac{1}{r} \sum_{A} w_{A}(\textbf{r}) = -\frac{1}{r}. 
\end{split}
\end{equation}
From Eqs.\ (\ref{eq:9}) and (\ref{eq:potential}), we have 
\begin{equation}
\label{eq:17a}
E^{\text{LFA}}_{x}[\rho_{\alpha},\rho_{\beta}] = \frac{1}{2} \sum_{\sigma=\alpha,\beta} \int \rho_{\sigma}(\textbf{r}) v^{\text{LFA}}_{x,\sigma}(\textbf{r}) d\textbf{r}, 
\end{equation} 
showing that the LFA exchange energy density per electron also has the correct $(-1/2r)$ asymptote \cite{LB94,March}. 

For the calculation of excitation energies using adiabatic linear-response TDDFT \cite{TDDFT1,TDDFT2,TDDFT3}, the functional derivative of $v^{\text{LFA}}_{x,\sigma}(\textbf{r})$ yields the LFA exchange kernel 
for $\sigma$-spin electrons, 
\begin{equation}
\label{eq:kernel} 
\begin{split}
f^{\text{LFA}}_{x,{\sigma}}(\textbf{r},\textbf{r}') & = \frac{\delta v^{\text{LFA}}_{x,\sigma}(\textbf{r})} {\delta\rho_{\sigma}({\bf r}')} 
= \frac{\delta^{2} E^{\text{LFA}}_{x}[\rho_{\alpha},\rho_{\beta}]}{\delta\rho_{\sigma}({\bf r}) \delta\rho_{\sigma}({\bf r}')} \\
& = -\frac{\text{erf}(\omega \left|{\bf r}-{\bf r}'\right|)}{\left|{\bf r}-{\bf r}'\right|} \sum_{A} \frac{w_{A}(\textbf{r})w_{A}(\textbf{r}')}{N_{A,\sigma}}. 
\end{split}
\end{equation}
In contrast to semilocal approximations, $f^{\text{LFA}}_{x,{\sigma}}(\textbf{r},\textbf{r}')$ in reciprocal space has the correct long-wavelength $O(-1/q^2)$ divergence as $q \rightarrow 0$, which is crucially important 
for the proper description of excitonic effects in non-metallic solids \cite{sol1,sol2,sol3,sol4}. We emphasize that this striking feature appears naturally from our fully nonlocal $E^{\text{LFA}}_{x}[\rho_{\alpha},\rho_{\beta}]$. 

To improve its description of short-range XC effects, $E^{\text{LFA}}_{x}[\rho_{\alpha},\rho_{\beta}]$ is combined with a popular semilocal functional, PBE \cite{PBE}. However, this will produce a 
double-counting (DC) energy $E_{\text{DC}}$, which needs to be removed. As the significant fraction of $E^{\text{LFA}}_{x}[\rho_{\alpha},\rho_{\beta}]$ should be from the core regions of the atoms, we presuppose 
that $\rho_{A,\sigma}(\textbf{r})$ is {\it strictly localized} at ${\bf R}_{A}$ (the position of the atom $A$) in Eq.\ (\ref{eq:9}), which gives $\rho_{A,\sigma}(\textbf{r}) \approx N_{A,\sigma}\delta(\textbf{r} - {\bf R}_{A})$ 
for satisfying Eq.\ (\ref{eq:NA}), to estimate $E_{\text{DC}}$, 
\begin{equation}
\label{eq:8b}
\begin{split}
E_{\text{DC}} =&\; -\sum_{\sigma=\alpha,\beta}\sum_{A} \frac{1}{2N_{A,\sigma}} \\ 
& \times \iint \bigg\lbrace N_{A,\sigma}\delta(\textbf{r} - {\bf R}_{A})\bigg\rbrace \bigg\lbrace N_{A,\sigma}\delta({\bf r}' - {\bf R}_{A})\bigg\rbrace \\
& \times \frac{\text{erf}(\omega \left|{\bf r} - {\bf r}' \right|)}{\left|{\bf r} - {\bf r}' \right|} d\textbf{r}d\textbf{r}' \\
=&\; -\sum_{\sigma=\alpha,\beta} \sum_{A} \frac{1}{2N_{A,\sigma}} (N_{A,\sigma})^{2} \bigg\lbrace \lim_{{\bf r}\to{\bf R}_{A}} \frac{\text{erf}(\omega \left|{\bf r} - {\bf R}_{A} \right|)}{\left|{\bf r} - {\bf R}_{A} \right|} \bigg\rbrace \\ 
=&\; -\sum_{\sigma=\alpha,\beta} \sum_{A} \frac{N_{A,\sigma}}{2} \bigg\lbrace \frac{2\omega}{\sqrt{\pi}} \bigg\rbrace \\ 
=&\; -\frac{\omega}{\sqrt{\pi}} \sum_{\sigma=\alpha,\beta} \sum_{A} N_{A,\sigma} = -\frac{\omega}{\sqrt{\pi}}N. 
\end{split}
\end{equation} 
This estimate is very accurate for systems with highly localized charges (e.g., HCl), and less accurate for systems with delocalized charges (e.g., benzene) \cite{Sup}. Our resulting LFA-PBE functional is given by 
\begin{equation}
\label{eq:8}
E^{\text{LFA-PBE}}_{xc} = E^{\text{PBE}}_{xc} + E^{\text{LFA}}_{x} - E_{\text{DC}}. 
\end{equation}
Note that $v^{\text{DC}}_{x,\sigma}(\textbf{r}) = \delta E_{\text{DC}}/\delta \rho_{\sigma}({\bf r}) = - \omega/\sqrt{\pi}$ is simply a constant, which can be absorbed into the constant $v^{\text{LFA}}_{x,\sigma}(\infty)$. 
As the KS potential is only defined within an arbitrary constant, without loss of generality, we require the KS potential to vanish asymptotically, which sets $v^{\text{LFA}}_{x,\sigma}(\infty) + \omega/\sqrt{\pi} = 0$. 
Unlike the FA model, LFA-PBE is size-consistent for any system composed of atoms \cite{Sup}. Note that LFA-PBE (with $\omega = 0$) reduces to PBE.

\subsection{RILFA scheme}

For systems composed of many atoms, LFA-PBE can be computationally unfavorable due to the numerical integration of many Hartree-like potentials in $v^{\text{LFA}}_{x,\sigma}(\textbf{r})$. To resolve this 
computational bottleneck without loss of much accuracy, Eq.\ (\ref{eq:potential}) can be efficiently evaluated by the resolution-of-identity (RI) approximation \cite{RIidea,Jaffe1996}. Following Ref.\ \cite{Jaffe1996}, 
$\rho_{A,\sigma}({\bf r})$ is expanded with an auxiliary basis set $\{g_{p}(\mathbf{r})\}$, i.e., $\rho_{A,\sigma}({\bf r}) \approx \tilde{\rho}_{A,\sigma}({\bf r}) = \sum_{p}a_{p}g_{p}(\mathbf{r})$, 
where the expansion coefficients $\{a_p\}$ are given by Eq.\ (8) of Ref.\ \cite{Jaffe1996} (with $\rho({\bf r})$ being replaced by $\rho_{A,\sigma}({\bf r})$). The RILFA exchange potential is evaluated by 
\begin{equation}
\label{RILFApotential}
v^{\text{RILFA}}_{x,\sigma}(\textbf{r}) = -\sum_{A} \frac{w_{A}(\textbf{r})}{N_{A,\sigma}} \int \tilde{\rho}_{A,\sigma}(\textbf{r}') \frac{\text{erf}(\omega \left|{\bf r}-{\bf r}'\right|)}{\left|{\bf r}-{\bf r}'\right|}d\textbf{r}'. 
\end{equation}
From Eq.\ (10) of Ref.\ \cite{Jaffe1996}, the RILFA exchange energy is given by 
\begin{equation}
\label{RILFAenergy}
\begin{split}
E^{\text{RILFA}}_{x} & = \sum_{\sigma=\alpha,\beta}\sum_{A} \bigg\lbrace -\frac{1}{N_{A,\sigma}} \\
& \times \iint \rho_{A,\sigma}({\bf r}) \tilde{\rho}_{A,\sigma}({\bf r}') \frac{\text{erf}(\omega \left|{\bf r} - {\bf r}' \right|)}{\left|{\bf r} - {\bf r}' \right|} d\textbf{r}d\textbf{r}' \\
& + \frac{1}{2 N_{A,\sigma}} \iint \tilde{\rho}_{A,\sigma}({\bf r}) \tilde{\rho}_{A,\sigma}({\bf r}') \frac{\text{erf}(\omega \left|{\bf r} - {\bf r}' \right|)}{\left|{\bf r} - {\bf r}' \right|} d\textbf{r}d\textbf{r}' \bigg\rbrace. 
\end{split}
\end{equation}
For a sufficiently large $\{g_{p}(\mathbf{r})\}$, the RILFA scheme approaches to the LFA scheme. Here, RILFA-PBE is defined by Eq.\ (\ref{eq:8}), with $E^{\text{LFA}}_{x}$ being replaced by $E^{\text{RILFA}}_{x}$.

\subsection{LFAs scheme}

For very large systems, both the LFA and RILFA schemes may be impractical, compared to the efficient semilocal density functional approach. Aiming to retain the correct $(-1/r)$ asymptote with essentially no added 
computational cost relative to semilocal functionals, the {\it strict} localization of $\rho_{A,\sigma}(\textbf{r})$ at ${\bf R}_{A}$ (i.e., $\rho_{A,\sigma}(\textbf{r}) \approx N_{A,\sigma}\delta(\textbf{r} - {\bf R}_{A})$) is presupposed 
in Eq.\ (\ref{eq:potential}), to define the LFAs exchange potential, 
\begin{equation} 
\label{eq:potential2} 
\begin{split}
v^{\text{LFAs}}_{x,\sigma}(\textbf{r}) & = -\sum_{A} \frac{w_{A}(\textbf{r})}{N_{A,\sigma}} \\
&\times \int \bigg\lbrace N_{A,\sigma}\delta({\bf r}' - {\bf R}_{A})\bigg\rbrace \frac{\text{erf}(\omega \left|{\bf r}-{\bf r}'\right|)}{\left|{\bf r}-{\bf r}'\right|}d\textbf{r}' \\
& = -\sum_{A}  w_{A}(\textbf{r}) \frac{\text{erf}(\omega \left|{\bf r}-{\bf R}_{A}\right|)}{\left|{\bf r}-{\bf R}_{A}\right|}. 
\end{split}
\end{equation} 
The asymptote of $v^{\text{LFAs}}_{x,\sigma}(\textbf{r})$ remains correct, 
\begin{equation}
\label{eq:potential3}
\lim_{r \rightarrow \infty} v^{\text{LFAs}}_{x,\sigma}(\textbf{r}) = -\sum_{A} w_{A}(\textbf{r}) \frac{1}{\left|{\bf r}\right|} = -\frac{1}{r}. 
\end{equation}
Based on Eq.\ (\ref{eq:17a}), the LFAs exchange energy is given by 
\begin{equation}
\label{eq:energy2}
E^{\text{LFAs}}_{x} = \frac{1}{2} \sum_{\sigma=\alpha,\beta} \int \rho_{\sigma}(\textbf{r}) v^{\text{LFAs}}_{x,\sigma}(\textbf{r})d\textbf{r}, 
\end{equation} 
to retain the correct $(-1/2r)$ asymptote of the LFAs exchange energy density per electron. Although $v^{\text{LFAs}}_{x,\sigma}(\textbf{r})$ differs from the functional derivative of $E^{\text{LFAs}}_{x}$ by a factor of 2, 
the prescribed LFAs scheme approaches to the LFA scheme for a sufficiently small $\omega$ value, where $v^{\text{LFAs}}_{x,\sigma}(\textbf{r})$ becomes an excellent approximation of $v^{\text{LFA}}_{x,\sigma}(\textbf{r})$. 
Similarly, LFAs-PBE is defined by Eq.\ (\ref{eq:8}), with $E^{\text{LFA}}_{x}$ being replaced by $E^{\text{LFAs}}_{x}$.

\section{Definition of an Optimal $\omega$ Value}

For the exact KS-DFT, the minus HOMO energy of a molecule should be the same as the vertical ionization potential (IP) of the molecule \cite{Fractional1,Fractional2,Fractional3,exactip1,exactip2}. 
Therefore, the optimal $\omega$ values 
for LFA-PBE, RILFA-PBE, and LFAs-PBE are determined by fitting the predicted IPs (calculated by the minus HOMO energies) of 18 atoms and 113 molecules in the IP131 database to the corresponding experimental IPs \cite{IP131}. 
All calculations are performed with a development version of \textsf{Q-Chem 3.2} \cite{Q-Chem}, using the 6-311++G(3df,3pd) basis set (and sufficiently large auxiliary basis sets for the RILFA scheme), unless noted otherwise. 
The error for each entry is defined as (error = theoretical value $-$ reference value). 

\begin{figure}[htbp] 
\includegraphics[scale=0.45]{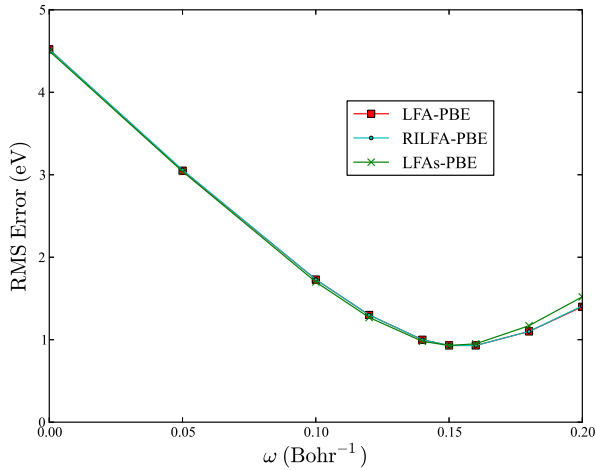} 
\caption{\label{fig:RMS} 
The root-mean-square (RMS) errors of LFA-PBE, RILFA-PBE, and LFAs-PBE for the IP131 database \cite{IP131}. The $\omega = 0$ case corresponds to PBE.} 
\end{figure} 

\begin{figure}[htbp] 
\includegraphics[scale=0.45]{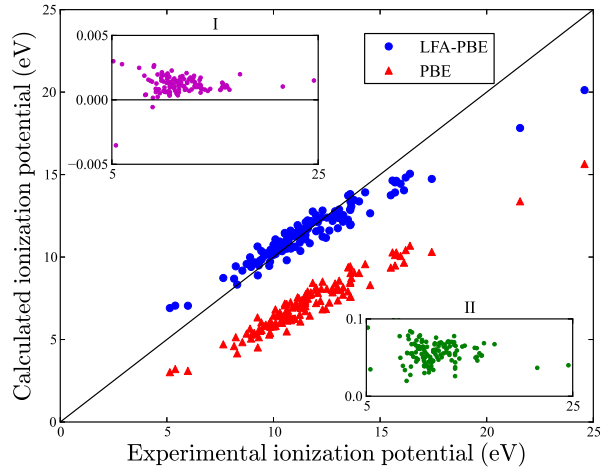} 
\caption{\label{fig:LFA-PBE_cal_exp} 
Calculated versus experimental ionization potentials (IPs) for the IP131 database \cite{IP131}. Inset I shows the differences between the IPs calculated by RILFA-PBE and LFA-PBE, while inset II shows 
the differences between the IPs calculated by LFAs-PBE and LFA-PBE. $\omega = 0.15$ Bohr$^{-1}$ is adopted for all the LFA-corrected PBE functionals.} 
\end{figure} 

As shown in Fig.\ \ref{fig:RMS}, the minimum root-mean-square (RMS) errors of LFA-PBE, RILFA-PBE, and LFAs-PBE for the IP131 database, which all occur at $\omega = 0.15$ Bohr$^{-1}$, are more than three times smaller 
than the RMS error of PBE (the $\omega = 0$ case), reflecting the importance of the correct asymptote of XC potential here \cite{Sup}. 

Adopting $\omega = 0.15$ Bohr$^{-1}$ for all the LFA-corrected PBE functionals, the calculated IPs are plotted against the experimental values in Fig.\ \ref{fig:LFA-PBE_cal_exp}. As can be seen, the differences between the IPs 
calculated by RILFA-PBE and LFA-PBE are within 0.005 eV, and the differences between the IPs calculated by LFAs-PBE and LFA-PBE are within 0.1 eV. Therefore, all the LFA-corrected PBE functionals yield very similar 
results, indicating that their XC potentials should be very similar \cite{Sup}. By contrast, the IPs calculated by PBE are seriously underestimated due to its incorrect asymptote. 

Similar results are found, when our LFA-related schemes are combined with LDA \cite{LDAX,LDAC}. As both the LDA and PBE XC potentials decay exponentially in the asymptotic region, their predicted IPs are similar, 
requiring essentially the same corrections from the LFA-related schemes (i.e., with the same optimal $\omega$) \cite{Sup}. Therefore, $\omega = 0.15$ Bohr$^{-1}$ can be recommended as the optimal $\omega$ value, 
when the LFA-related schemes are combined with a local or semilocal functional whose functional derivative has the (incorrect) exponential asymptote.

\section{Results and Discussion}

Here, we examine the performance of the PBE and LFA-corrected PBE functionals ($\omega = 0.15$ Bohr$^{-1}$) on various test sets, involving 
the reaction energies of 30 chemical reactions (a test set described in Ref.\ \cite{wB97X}), the 223 atomization energies (AEs) of the G3/99 set \cite{G2a1,G2a2,G2a3}, 
the 76 barrier heights (BHs) of the NHTBH38/04 and HTBH38/04 sets \cite{BH1,BH2}, the 22 noncovalent interactions of the S22 set \cite{S22}, 19 valence excitation energies, and 23 Rydberg excitation energies. 
There are in total 393 pieces of data in our test sets, which are quite large and diverse. Unspecified detailed information of the test sets is given in Ref.\ \cite{IP131}. 
For comparison, the results calculated by the LB94 potential (a popular AC model XC potential) \cite{LB94} are taken from Ref.\ \cite{LCAC}. 
Note that the LB94 potential is a linear combination of the LDA exchange potential, the LDA correlation potential, and a gradient-dependent exchange potential (e.g., see Eq.\ (55) of Ref.\ \cite{LB94}). 
Due to the inclusion of the gradient-dependent exchange potential, the LB94 potential is {\it not} a functional derivative \cite{Staroverov,Staroverov2b}. 
In Ref.\ \cite{LCAC}, the exchange energy from the LB94 exchange potential was evaluated by the popular Levy-Perdew virial relation \cite{virial} (e.g., see Eq.\ (1) of Ref.\ \cite{LCAC}), 
while the correlation energy from the LB94 correlation potential was directly evaluated by the LDA correlation energy functional. 

As shown in Table \ref{tab:property}, the performance of the LFA-corrected PBE functionals is similar to that of PBE \cite{Sup}. As these properties are rather insensitive to the asymptote of the XC potential, our schemes 
do not affect the already good performance of PBE. By contrast, due to the lack of $E_{xc}[\rho]$, LB94 performs the worst. Therefore, one should avoid using the AC model potential scheme for the calculation of 
total energies and related properties. 

For the valence and Rydberg excitation energies, we perform adiabatic linear-response TDDFT calculations, using the 6-311(2+,2+)G** basis set, on five molecules: nitrogen gas (N$_2$), carbon monoxide (CO), 
water (H$_2$O), ethylene (C$_2$H$_4$), and formaldehyde (CH$_2$O) on the experimental geometries taken from Ref.\ \cite{ver_ref}. For the TDDFT calculations using the LFA-corrected PBE functionals, both 
the PBE XC kernel and the LFA exchange kernel should be adopted for a consistent approximation on $E_{xc}[\rho]$. However, in this work, we only adopt the PBE XC kernel, and neglect the LFA exchange kernel 
for computational simplicity. Note that the similar tricks have been constantly used in the AC model potential approach (e.g., the XC kernel of a local or semilocal functional is 
adopted) \cite{AC0,AC2,LB94_ALDA0,LB94_ALDA,LB94_ALDA2,LB94_ALDA3,LCAC}. For example, the LDA XC kernel is frequently adopted for the TDDFT calculations using the LB94 potential \cite{LCAC}. 
For finite systems, this approximation should not make much difference in the prediction of valence and Rydberg excitation energies. 
As shown in Table \ref{tab:VRes}, all the LFA-corrected PBE functionals and LB94 perform well for both the valence and Rydberg excitations, while PBE severely underestimates Rydberg excitation energies due to 
its incorrect asymptote \cite{Sup}. 

\begin{table} 
\begin{ruledtabular} 
\caption{\label{tab:property} Mean absolute errors (in kcal/mol) of various test sets (see the text for details). 
The LB94 results are taken from Ref.\ \cite{LCAC}. (1 kcal/mol = 0.0434 eV.)} 
\begin{tabular}{lccccc}
System & PBE & LFA-PBE & RILFA-PBE & LFAs-PBE & LB94 \\ 
\hline
Reaction (30) &    4.38 &   4.48  &  4.47  &  4.42  &               \\ 
G3/99 (223)    &  21.51 & 27.36 & 27.36 & 24.58 & 484.91 \\
NHTBH (38)   &     8.62 &   8.71 &   8.71 & 8.66   & 93.94 \\
HTBH (38)      &     9.67 &   9.69 &   9.69 & 9.69   & 44.31 \\
S22 (22)          &     2.72 &   2.37 &   2.37 & 2.52   & 51.70 \\
\end{tabular}
\end{ruledtabular}
\end{table}

\begin{table}
\begin{ruledtabular}
\caption{\label{tab:VRes} Mean absolute errors (in eV) of the 19 valence and 23 Rydberg excitation energies of five molecules \cite{ver_ref}. 
The LB94 results are taken from Ref.\ \cite{LCAC}.} 
\begin{tabular}{lcccccc}
System & PBE & LFA-PBE & RILFA-PBE & LFAs-PBE & LB94 \\ 
\hline
Valence (19)  &  0.32  &  0.29  &  0.29  &  0.29 & 0.36 \\ 
Rydberg (23) &  1.30  &  0.46  &  0.46  &  0.49  & 0.73 \\ 
\end{tabular}
\end{ruledtabular}
\end{table}

\section{Conclusions}

In conclusion, we have developed the LFA scheme, wherein an exchange density functional whose functional derivative has the correct $(-1/r)$ asymptote can be directly added to any semilocal density functional. 
In contrast to semilocal approximations, the LFA exchange kernel in reciprocal space exhibits the desirable singularity of the type $O(-1/q^2)$, which is an important feature for the description of excitonic effects in non-metallic 
solids. Applying the LFA scheme to PBE, the resulting LFA-PBE ($\omega = 0.15$ Bohr$^{-1}$) has yielded accurate IPs and Rydberg excitation energies for a wide range of atoms and molecules, while performing similarly 
to PBE for various properties that are insensitive to the asymptote. Without loss of accuracy, two closely related schemes (RILFA and LFAs) have been developed for the efficient treatment of large systems. 
Relative to the popular model XC potential scheme, LFA-PBE is significantly superior for ground-state energies and related properties. It remains to be seen if the LFA-corrected PBE functionals will perform well 
for properties sensitive to the details of the XC potential (not just to the asymptote), such as quantum defects \cite{QDef}. 

As with all pure density functional methods (e.g., semilocal functionals and model XC potentials), some limitations remain. Due to the lack of HF exchange, the LFA-corrected PBE functionals may suffer from the SIE problems, 
energy-gap problems, and charge-transfer problems (e.g., see the discussions in Ref.\ \cite{LCAC}). Nevertheless, the energy-gap problems may be circumvented by the perturbation approach recently developed 
in Ref.\ \cite{DDChai}. 
Although the LC hybrid scheme, which has remedied several qualitative failures of pure density functional methods, could be reliably accurate for a very wide range of applications \cite{LCAC}, it can be impractical 
for very large systems due to the expensive computational cost. By contrast, our LFAs-PBE, which has the correct $(-1/r)$ asymptote with essentially no added computational cost relative to PBE, is potentially very useful 
for the study of the ground-state energies and related properties, frontier orbital energies, valence and Rydberg excitation energies, and time-dependent properties of very large systems.

\begin{acknowledgments}

This work was supported by the National Science Council of Taiwan (Grant No. NSC101-2112-M-002-017-MY3), National Taiwan University (Grant Nos.\ 99R70304, 101R891401, and 101R891403), 
and the National Center for Theoretical Sciences of Taiwan. We thank Dr. Yihan Shao (Q-Chem, Inc.) for helpful discussions. 

\end{acknowledgments}

\bibliographystyle{pra}

\end{document}